\documentclass[showkeys,showpacs,preprintnumbers,aps,twocolumn,floatfix]{revtex4}
\usepackage{bm}
\usepackage[english]{babel}
\usepackage[latin1]{inputenc}
\usepackage{graphicx}
\usepackage{amsmath,amssymb}

\begin{document}

\date{\today}
\title{Controlling surface morphologies by time-delayed feedback}
\author{M.~Block$^{1}$}
\email{block@itp.physik.tu-berlin.de}
\thanks{Fax: +49-(0)30-314-21130}
\author{B.~Schmittmann$^{2}$}
\author{E.~Sch{\"o}ll$^{1}$}
\affiliation{$1$ Institut f{\"u}r Theoretische Physik, Technische Universit{\"a}t Berlin,
D-10623 Berlin, Germany\\
$2$ Department of Physics, Virginia Tech, Blacksburg, VA 24061, USA}

\begin{abstract}
We propose a new method to control the roughness of a growing surface, via a
time-delayed feedback scheme. As an illustration, we apply this method to
the Kardar-Parisi-Zhang equation in 1+1 dimensions and show that the
effective growth exponent of the surface width can be stabilized at any
desired value in the interval $[0.25,0.33]$, for a significant length of
time. The method is quite general and can be applied to a wide range of
growth phenomena. A possible experimental realization is suggested.
\end{abstract}

\maketitle

\parindent = 0pt

%%%%%%%%%%         Abstract            %%%%%%%%%%%%%%

%%%%%%%%%%        Introduction          %%%%%%%%%%%%%%
\emph{Introduction.} The control of unstable states in chaotic or
pattern-forming nonlinear dynamic systems has attracted much interest
recently \cite{SCH99,NIJ96}. Time-delayed feedback control \cite{PYR92} has
been especially successful in stabilizing a variety of dynamic and spatial
structures, including noise-induced oscillations and patterns found, e.g.,
in semiconductor nanostructures \cite{SCH01,SCH04,STE05a,SCH06a}. The
fabrication of such nanostructures typically involves the deposition of a
material onto a substrate. One of the primary experimental goals is to
achieve nanoscale control of layer thickness and surface (or interface)
morphology. On the theoretical side, considerable effort has focused on
developing suitable evolution equations for the growing layer and its
surface \cite{KRU97}. While many different versions 
\cite{LAI91,KRU91,NAT92,KRU93,BAR95,DAS96,MAR96,SPJ06} of these equations exist,
depending on the details of deposition processes and molecular interactions
and kinetics, all of them share certain fundamental characteristics: they
are noisy, nonlinear partial differential equations in space and time, and
describe an important class of generic nonequilibrium phenomena.

It is natural to ask whether the control techniques of nonlinear dynamics
can be successfully applied to surface growth problems. The goal is, of
course, to stabilize desired surface characteristics, such as its
spatio-temporal height-height correlations or its roughness during the
growth process. Even if such control can only be achieved in a finite window
of time, its experimental potential is undiminished since the deposition
process can simply be terminated at the desired time, thanks to today's
precise in situ characterization capabilities. In this letter, we provide a
first set of answers to this question. We choose the most promising type of
control, time-delayed feedback, and study its effects on a paradigmatic
growth model, the Kardar-Parisi-Zhang (KPZ) equation \cite{KAR86}.
Specifically, we attempt to control the \emph{effective }dynamic growth
exponent $\beta $ associated with the roughness of the growing surface.
Implementing two realizations of the control scheme, we will see below that
we can indeed stabilize $\beta $ in a range of values between the two
universal limits, $1/4$ and $1/3$, over at least one to two decades in time.

This letter is organized as follows. We first introduce the KPZ equation and
our numerical solution scheme, accompanied by a representative data set for
the growing surface roughness, and specifically its growth exponent, in the
absence of control. Next, we implement two types of time-delayed feedback
control and demonstrate how reasonably accurate values of the (effective)
growth exponent can be achieved. We conclude with a summary and a discussion
of open questions.

\emph{The KPZ equation.} If overhangs and bulk fluctuations can be
neglected, growth phenomena are often modelled in terms of nonlinear
stochastic partial differential equations. A single-valued variable, 
$h(x,t)$, denotes the height of the surface above a reference 
plane and fluctuates as a function of time $t$ and 
position $x$ (measured in this d-dimensional plane). 
The simplest such equation is the KPZ equation \cite{KAR86} which
describes the growth of a surface in the absence of any conservation laws: 
\begin{equation}
\partial _{t}h(x,t)=\nu \nabla ^{2}h(x,t)+\frac{\lambda }{2}(\nabla
h)^{2}+\eta (x,t)  \label{eq:kpz}
\end{equation}%
Here, $\nu >0$ denotes an interface smoothing term, associated with a
surface tension; the nonlinear coupling $\lambda $ reflects the strength of
lateral growth, and $\eta (x,t)$ models the height fluctuations due to
random deposition of material. An overall, constant growth velocity has
already been eliminated, by transforming into a suitable co-moving frame.
Focusing on large-scale, long-time properties of the surface, it is
sufficient to consider Gaussian white noise, i.e., 
\begin{eqnarray}
<\eta (x,t)> &=&0  \notag \\
<\eta (x,t)\eta (x^{\prime },t^{\prime })> &=&2D\delta ^{d}(x-x^{\prime
})\delta (t-t^{\prime })  \label{eq:noise}
\end{eqnarray}%
The KPZ equation has been discussed in many different contexts, including
thin film growth \cite{OJE00,OJE03,AUG06,SUR99}, fluctuating hydrodynamics %
\cite{FOR77}, driven diffusive systems \cite{BEI85,JAN86}, tumor growth in
biophysics \cite{BRU03,BRU98}, propagating fire fronts \cite{MAU99,MAU97}
and econophysics \cite{BAL04}. In the following, we will use the language of
surface growth, but our findings are easily translated into these other
contexts and just as relevant there.

For simplicity, we restrict ourselves to one spatial dimension. We monitor
the time-dependence of the (root mean square) surface roughness $w$, defined
by 
\begin{equation}
w^{2}(L,t)=\frac{1}{L}\left\langle \sum_{x}[h(x,t)-\bar{h}%
(t)]^{2}\right\rangle   \label{eq:def_w}
\end{equation}%
Here, $L$ denotes the system size, and $\bar{h}(t)\equiv L^{-1}\sum_{x}h(x,t)
$ is the mean surface height at time $t$. Configurational averages are
denoted by $\left\langle ...\right\rangle $. The sum over $x$ anticipates
the space discretization associated with the numerical integration scheme.
It is well known that $w\,$ obeys scaling in the form $w(L,t)=L^{\alpha
}f(t/L^{z})$ where $f$ is a scaling function and $\alpha $ and $z$ denote
the roughness and dynamic exponents, respectively \cite{FAM85}. In the
saturation regime $t/L^{z}\gg 1$, $f$ approaches a constant so that $w\sim
L^{\alpha }$ becomes independent of time. In contrast, in the growth regime $%
t/L^{z}\ll 1$, the width grows as a power of time, $w\sim t^{\beta }$, with
a growth exponent $\beta $. Consistency with the general scaling form
imposes an exponent identity, $\beta =\alpha /z$. The values for the
exponents $\alpha $, $\beta $, and $z$ are universal. Two universality
classes can be distinguished. If the nonlinear term of the KPZ equation
vanishes ($\lambda =0$),\ the equation reduces to the exactly soluble
Edwards-Wilkinson (EW) equation \cite{EDW82}, with $\alpha =1/2$ and $z=2$
whence $\beta =1/4$. In contrast, any \emph{nonzero} value of $\lambda $
belongs to the KPZ universality class with $\alpha =1/2$ and $z=3/2$ whence $%
\beta =1/3$ \cite{KAR86}. These values give us some benchmarks against which
we can check our numerical scheme. We use a forward-backward Euler method %
\cite{LAM98a} to solve the stochastic differential equations numerically.
The associated discretization captures certain prefactors much more
accurately than the standard (naive) scheme \cite{LAM98}. Our integration
parameters are $\nu =0.1$ and $D=0.5$; the space discretization is set to $%
\Delta x=1$, with $L=1024$ and $4096$, and the time increment is set at $%
\Delta t=10^{-3}$. Finally, $\lambda $ varies between $0.00$ and $0.25$. The
upper cutoff is chosen so as to avoid numerical instabilities. 

%=============================
\begin{figure}[h]
\begin{center}
\includegraphics[width=.48\textwidth]{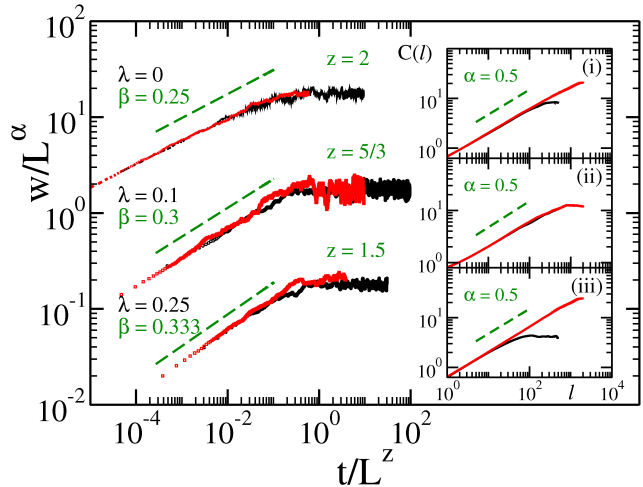}
\end{center}
\caption{(color online). Scaling plots of the roughness $w(L,t)$ and height-height
correlation function $C(l)\equiv \left \langle h(l,t) h(0,t) \right \rangle \sim l^{\alpha}$ (insets) for system sizes $L=1024$ (red) and $L=4096$ (black) and three
choices of $\lambda$: (i) $\protect%
\lambda =0$ (shifted by a factor 100), (ii) $\protect\lambda =0.1$
(shifted by a factor 10), (iii) $\protect\lambda =0.25$. $\protect\nu =0.1$ for all data sets. The green broken lines provide guides to the eye.}
\label{fig:uncontrolled solutions}
\end{figure}
%================================

Figure \ref{fig:uncontrolled solutions} shows a scaling plot for $w(L,t)$
before any control schemes are implemented. Data for three different values
of $\lambda $ are shown. The roughness exponent $\alpha $ is consistent with 
$1/2$, independent of $\lambda $, as expected. For $\lambda =0$, we see
excellent data collapse with the EW scaling exponents, and the KPZ exponents 
are confirmed for the largest $\lambda$. The latter should be universal, 
for all $\lambda \neq 0$; however, for $0 < \lambda < 0.25$,  
strong crossover effects between EW and KPZ behavior are observed.
Remarkably, this crossover manifests itself as a surprisingly
clean power law, with a $\lambda $-dependent $\emph{effective}$ growth
exponent $\beta$ \emph{below} $1/3$. Eventually, the asymptotic value ($1/3$) is
reached, but only after an $L$- and $\lambda $-dependent crossover time.  

\emph{Systems with control.} We now turn to possible control mechanisms. For
chemical vapor deposition of silica films, there is some experimental
evidence \cite{OJE00} that the lateral growth velocity is related to the
temperature, via a temperature-depending sticking probability. In other
words, $\lambda $ -- and hence effective growth exponents -- can be
controlled via the temperature. For our differential equation, we tune $%
\lambda $ directly, in order to stabilize a \emph{desired} effective growth
exponent, $\beta _{0}$. In detail, the scheme is as follows. First, we
choose the desired value of the growth exponent, $\beta _{0}$, and select an
appropriate time delay $\tau $. Generating sufficiently many samples of $%
h(x,t)$, we record $w(t-\tau )$ and $w(t)$ (the argument $L$ will be omitted
from \ now on). The \emph{local} exponent $\beta _{local}$ at time $t$ is
defined as 
\begin{equation}
\beta _{local}(t)\equiv \frac{\log w(t)-\log w(t-\tau )}{\log t-\log (t-\tau
)}  \label{eq:local growth exponent}
\end{equation}%
Depending on the sign and value of $\beta _{local}(t)-\beta _{0}$, we adjust
the nonlinear coupling, $\lambda $, of the KPZ equation, as follows. First,
we introduce a control function $F(t)$. For \emph{digital} control, we
define 
\begin{equation}
F(t)\equiv 
\begin{cases}
a,\; & if\;\beta _{local}\leq \beta _{0}\cr-a,\; & if\;\beta _{local}>\beta
_{0}%
\end{cases}
\label{eq:control function digital control}
\end{equation}%
where the parameter $a$ defines the control ``bit'', i.e., the amount by
which $\lambda $ changes at each control step. Alternatively, we also
investigate a \emph{differential} method for which 
\begin{equation}
F(t)\equiv K(\beta _{0}-\beta _{local})
\label{eq:control function differential control}
\end{equation}%
and $K$ sets the amplitude of the control strength. Given one of the two
choices of $F(t)$, the control scheme sets in at time $t_{0}$. From then on,
the nonlinearity $\lambda $ is updated at times $t_{n}\equiv t_{0}+n\tau $, $%
n=1,2,...$, starting from an initial value $\lambda _{0}$, according to 
\newline
\begin{equation}
\lambda (t)=%
\begin{cases}
\lambda _{0}, & if\;t<t_{0}\cr\lambda (t-\tau )+F(t), & if\;t=t_{n}\cr%
\lambda (t_{n}), & if\;t_{n}<t<t_{n+1}\cr%
\end{cases}
\label{eq:control}
\end{equation}%
Our scheme is successful if $\beta _{local}(t)$ approaches $\beta _{0}$ and
then settles at the desired value within a reasonable time frame after the
control has been activated.

Some comments are in order.
Starting from a random initial condition, we first choose a starting value, 
$\lambda _{0}$, for the nonlinearity and integrate the KPZ equation without
control up to time $t_{0}$, in order to eliminate transients. A reasonably
stable growth regime is achieved around $t_{0}\sim 10$, independent of $L$ 
(provided $L$ is not too small, i.e., $L \geq 64$).
Then, we turn on the
control, following either the digital or the differential scenario.
Regarding the choice of the time delay $\tau $, it must be large enough
compared to the time increment $\Delta t$ so as not to interfere with the
integration procedure, but small enough to provide responsive control. We
find that we get good results for a time delay $0.1<\tau <1.0$. Similarly,
we choose the control amplitudes $a$ and $K$ such that the increments in 
$\lambda $ are small compared to $\lambda _{0}$ but large enough to generate
a noticeable response. For example for $\tau = 1.0$, choosing $a$ in 
the range $[0.002,0.02]$ and $K$ in the range $[0.005,0.05]$ provides the 
best results. 

%==============================
\begin{figure}[tbp]
\begin{center}
\includegraphics[width=.45\textwidth]{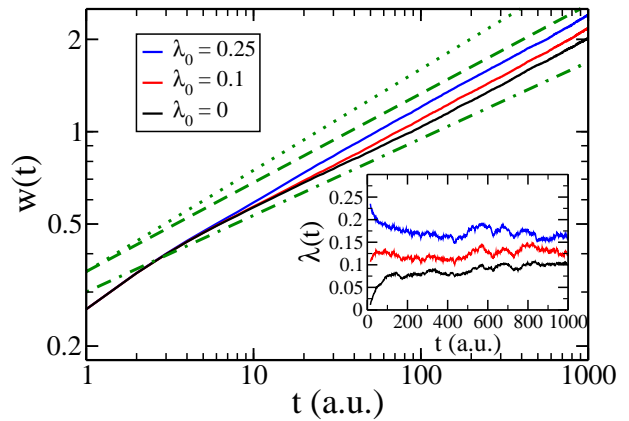}
\end{center}
\caption{(color online). Roughness and control function (inset) evolution for digital time
delayed feedback control for strong (blue), weak (red), and zero (black) 
initial nonlinearity $\protect\lambda _{0}$. The desired effective growth
exponent is set at $\beta_0=0.29$. To provide a comparison, the straight (green) 
lines have slopes $0.33$ (dotted), $0.29$ (dashed), and $0.25$ (dash-dotted).
All data sets are obtained with $\protect\nu = 0.1$ and $a = 0.01$.}
\label{fig:digital control}
\end{figure}
%===============================

\begin{figure}[t!]
\begin{center}
%\vspace{1.4cm} 
\includegraphics[width=.45\textwidth]{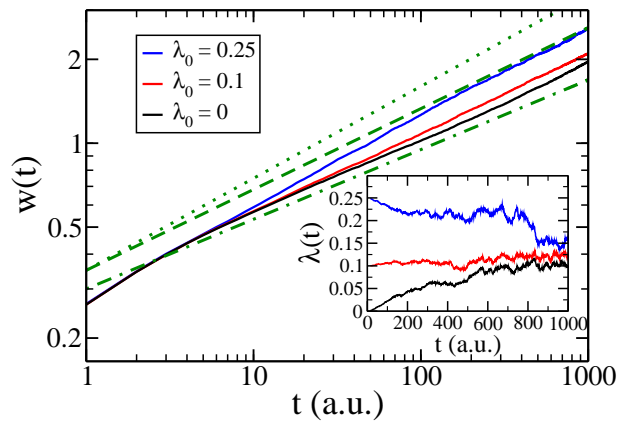}
\end{center}
\caption{(color online). Roughness and control function (inset) evolution for differential time
delayed feedback control for strong (blue), weak (red), and zero (black) 
initial nonlinearity $\protect\lambda _{0}$. The desired effective growth
exponent is set at $\beta_0=0.29$. To provide a comparison, the straight (green) 
lines have slopes $0.33$ (dotted), $0.29$ (dashed), and $0.25$ (dash-dotted).
All data sets are obtained with $\protect\nu = 0.1$ and $K = 0.02$.}
\label{fig:differential control}
\end{figure}

\emph{Results}. Figures \ref{fig:digital control} and \ref{fig:differential control} 
show our results. Starting from three initial values of $\lambda_0$, namely, 
$0$, $0.1$, and $0.25$, we attempt to stabilize the effective growth exponent at 
$\beta_0=0.29$, mid-way between the KPZ and EW values. 
Irrespective of $\lambda_0$, we find that both digital and differential
control result in an effective growth exponent very close to the desired value, 
over at least a decade of integration time ($ 100\lesssim t \lesssim 1000$). 

%==========================================
%===================================

For sufficiently large time, the control function $\lambda(t)$ appears to approach 
a constant value, close to $0.15$. However, the details of the approach 
depend on the initial $\lambda_0$. 
For strong initial nonlinearity $\lambda _{0}=0.25$, $\lambda(t)$ is
approximately monotonically decreasing, apart from significant fluctuations. 
For both weak and vanishing initial coupling, $\lambda(t)$ approaches its 
``limit'' from below. This behavior is observed for both, digital and differential, 
control. We tested several other choices of $0.25\leq \beta _{0}<0.33$ and 
$\lambda_0$, and found similar behavior.

Experimentally, it is usually desirable to achieve small roughness. To
push our control schemes to the limit, we test the most extreme case,
namely, $\beta _{0}=0.25$ with large initial nonlinearity $\lambda _{0}=0.25$.
Before the control sets in, the roughness grows considerably faster than
$t^{0.25}$. As soon as the control sets in, $\lambda (t)$ decreases quite
dramatically, leading to a reduction of the effective $\beta$. However, over
the time period considered ($t \leq 1000$), it never decreases far enough 
to reach the desired $0.25$. 

Finally, we note that it is not possible to achieve exponent values 
\emph{outside} the interval $[0.25,0.33]$. Choosing $\beta _{0}>0.33$ generates
unbounded growth of the control function $\lambda (t)$, accompanied by
instabilities in the integration routine. Similarly, $\beta _{0}<0.25$
quickly leads to large \emph{negative }values of $\lambda (t)$ which tend to
favor KPZ exponents (since the sign of $\lambda $ plays no role). As a
result, $\lambda (t)$ becomes even more negative until a numerical
instability occurs. To avoid this instability, we also implemented a
symmetrized version of control (with $a\rightarrow -a$ when $\lambda (t)<0$%
). In this case, $\lambda (t)$ approaches zero and fluctuates about it, so
that the effective exponent settles at $0.25$.

\emph{Conclusions. }To summarize, both digital and differential control are
rather successful at stabilizing effective growth exponents in the KPZ
equation. For the relatively small system sizes used here, these exponents
can be tuned in the range $[0.25,0.33]$, i.e., within the limits set by the
EW and the KPZ equation, respectively. Let us emphasize again that only the
values $1/4$ (for $\lambda =0$) and $1/3$ (for any $\lambda \neq 0$)
correspond to true asymptotic exponents; for larger system sizes and longer
integration times, these emerge clearly. However, for small systems, we
observe surprisingly clean effective growth exponents which appear to depend
monotonically on the magnitude of the nonlinearity. Hence, it is possible to
choose a desired reference exponent $\beta _{0}$ and implement a
time-delayed control of the nonlinearity in such a way that the effective
exponent first approaches $\beta _{0}$ and then stabilizes at that value for
a significant length of time (roughly, $10^{2}\lesssim t\lesssim 10^{3}$, in
our units). In all simulations, the (stationary) roughness remains constant
at $\alpha =0.5$, reflecting the value $1/2$ which is common to both the EW
and the KPZ equation. The control protocol itself is independent from the
dimension of the surface. Work is in progress to test other growth equations
and to extend the KPZ study to 2+1 dimensions. We thus hope to develop a
control tool which might also be useful in experimental setups. The work of
Ojeda et al \cite{OJE00,OJE03} gives some indications that the KPZ
nonlinearity is tunable via the temperature. Hence, it seems feasable to
implement a time-delayed feedback loop and stabilize desired growth
exponents by suitable adjustments of the temperature.

\begin{acknowledgments}
We have benefitted from helpful discussions with Uwe T{\"a}uber and
Erwin Frey. BS wishes to thank the SFB 296 and the ITP at TU Berlin 
for their hospitality.
This work was supported in part by the Deutsche Forschungsgemeinschaft 
through SFB 296, and by the U.S. National Science Foundation 
through DMR-0414122.
\end{acknowledgments}

%\bibliographystyle{prsty-fullauthor}
%\bibliography{blo06}

\end{document}